\documentclass[sigconf]{acmart}

\def\BibTeX{{\rm B\kern-.05em{\sc i\kern-.025em b}\kern-.08emT\kern-.1667em\lower.7ex\hbox{E}\kern-.125emX}}
    
\usepackage{balance}
\begin{document}

%
\title[Mapping Informal Settlements in Developing Nations]{Mapping Informal Settlements in Developing Countries using Machine Learning and Low Resolution Multi-spectral Data}

\author{Bradley J Gram-Hansen}
\authornote{Both authors contributed equally to this research.}
\affiliation{
  \institution{University of Oxford}
  \city{Oxford}
  \country{UK}
}
\email{bradley@robots.ox.ac.uk}

\author{Patrick Helber}
\authornotemark[1]
\affiliation{%
  \institution{DFKI, TU Kaiserslautern }
  \city{Kaiserslautern}
  \country{Germany}
  }
\email{patrick.helber@dfki.de}

\author{Indhu Varatharajan}
\affiliation{%
  \institution{DLR Institute for Planetary}
  \city{Berlin}
  \country{Germany}
}
\email{indhu.varatharajan@dlr.de}

\author{Faiza Azam}
\affiliation{%
 \institution{Independent Researcher}
}
 
\author{Alejandro Coca-Castro }
\affiliation{%
  \institution{Kings College London}
  \city{London}
  \country{UK}
  }
  \email{alejandro.coca_castro@kcl.ac.uk}

\author{Veronika Kopackova}
\affiliation{%
  \institution{Czech Geological Survey}
  \city{Prague}
  \country{Czech Republic}
}
\email{Veronika.Kopackova@seznam.cz}

\author{Piotr Bilinski}
\affiliation{\hspace{30px}\makebox[0pt]{\institution{University of Oxford  \& University of Warsaw}}\newline
\city{Oxford}
\country{UK}
}
\email{piotrb@robots.ox.ac.uk}
%
\renewcommand{\shortauthors}{Gram-Hansen and Helber, et al.}
%
\begin{abstract}
Informal settlements are home to the most socially and economically vulnerable people on the planet. In order to deliver effective economic and social aid, non-government organizations~(NGOs), such as the United Nations Children's Fund (UNICEF), require detailed maps of the locations of informal settlements. However, data regarding informal and formal settlements is primarily unavailable and if available is often incomplete. This is due, in part, to the cost and complexity of gathering data on a large scale.  
To address these challenges, we, in this work, provide three contributions. 
1) A brand new machine learning data-set, purposely developed for informal settlement detection.
2) We show that it is possible to detect informal settlements using freely available low-resolution~(LR) data, in contrast to previous studies that use very-high resolution~(VHR) satellite and aerial imagery, something that is cost-prohibitive for NGOs. 3) We demonstrate two effective classification schemes on our curated data set, one that is cost-efficient for NGOs and another that is cost-prohibitive for NGOs, but has additional utility. We integrate these schemes into a semi-automated pipeline that converts either a LR or VHR satellite image into a binary map that encodes the locations of informal settlements.
\end{abstract}

\copyrightyear{2019} 
\acmYear{2019} 
\setcopyright{acmlicensed}
\acmConference[AIES '19]{AAAI/ACM Conference on AI, Ethics, and Society}{January 27--28, 2019}{Honolulu, HI, USA}
\acmBooktitle{AAAI/ACM Conference on AI, Ethics, and Society (AIES '19), January 27--28, 2019, Honolulu, HI, USA}
\acmPrice{15.00}
\acmDOI{10.1145/3306618.3314253}
\acmISBN{978-1-4503-6324-2/19/01}
\settopmatter{printacmref=true}
\fancyhead{}
%

%

%
\begin{teaserfigure}
\centering
\includegraphics[width=\textwidth, height=3.5cm, keepaspectratio]{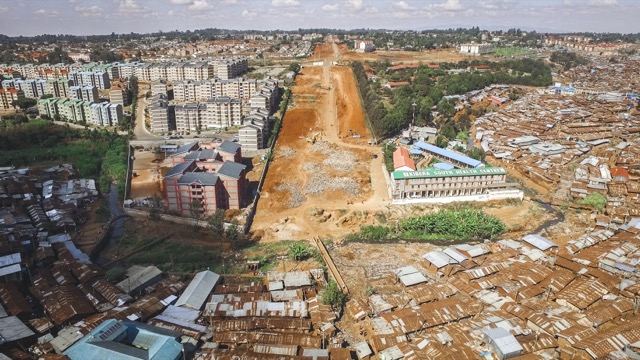}
\caption{Divide between formal and informal settlements in Kibera, Nairobi. Copyright Unequal Scenes and Johnny Miller. }
  \Description{An example of informal settlements.}
  \label{fig:teaser}
\end{teaserfigure}

\maketitle

\section{Introduction}

The United Nations~(UN) state that inhabitants of settlements that meet any of the following criteria are defined to be living in an informal settlement~\cite{united2012state}: 
\begin{quote}
\begin{enumerate}
\item Inhabitants have no security of tenure vis-\`a-vis the land or
dwellings they inhabit, with modalities ranging from squatting to informal rental housing. 
\item The neighborhoods usually lack, or are cut off from, basic services and city infrastructure.
\item The housing may not comply with current planning and building regulations, and is often situated in geographically and environmentally hazardous areas.
\end{enumerate}
\end{quote}
Slums, an example of informal settlements, are the most deprived and excluded form of informal settlements. They can be characterized by poverty and large agglomerations of dilapidated housing, located in the most hazardous urban land, near industries and dump sites, in swamps, degraded soils and flood-prone zones~\cite{Kohli2016}. Slum dwellers are constantly exposed to eviction, disease and violence~\cite{Sclar2005}, which stems from and leads to more severe economic and social constraints~\cite{WEKESA2011238}. Although informal settlements are well studied in the humanities and remote sensing communities \cite{fincher2003planning,WEKESA2011238,united2012state,huchzermeyer2006informal,hofmann2008detecting} in machine learning, only a small amount of research has been conducted on informal settlements, with all of that research using VHR and high resolution(HR) satellite imagery~\cite{mahabir2018critical,mboga2017detection,varshney2015targeting}, a cost prohibitive option for many NGOs and governments of developing nations. In contrast, there is an abundance of freely available and globally accessible LR satellite imagery, provided by the European Space Agency~(ESA), which provides updated imagery of the entire land mass of the Earth every 5 days~\cite{Waitr2016,Esas2img,Esasent2}. To the authors knowledge, no previous approaches have used LR imagery.

The ability to map and locate these settlements would give organizations such as UNICEF and other NGOs the ability to provide effective social and economic aid~\cite{pais2002poverty}. This in turn would enable those communities to evolve in a sustainable way, allowing the people living in those environments to gain a much better quality of life addressing multiple of the UN sustainable development goals~\cite{UNsus}. These goals aim to eliminate poverty, increase good health and well-being, provide quality education, clean water and sanitation, affordable and clean energy, sustainable work and economic growth, access to industry, innovation and infrastructure.\\ 

\begin{figure}[t!]
\begin{minipage}{.25\textwidth}
  \includegraphics[width=0.8775\linewidth]{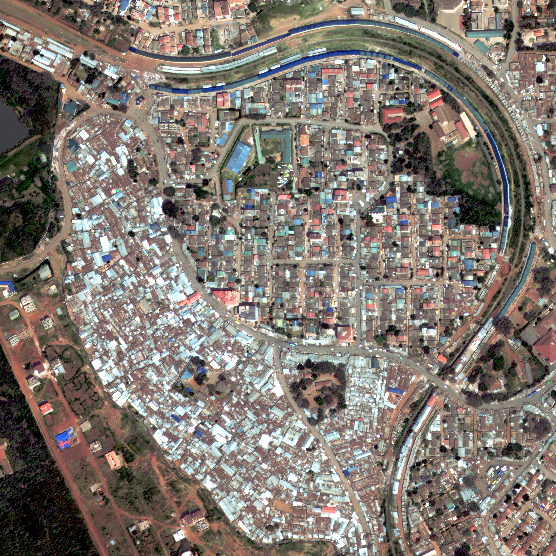}
\end{minipage}%
\begin{minipage}{.25\textwidth}
  \includegraphics[width=0.8775\linewidth]{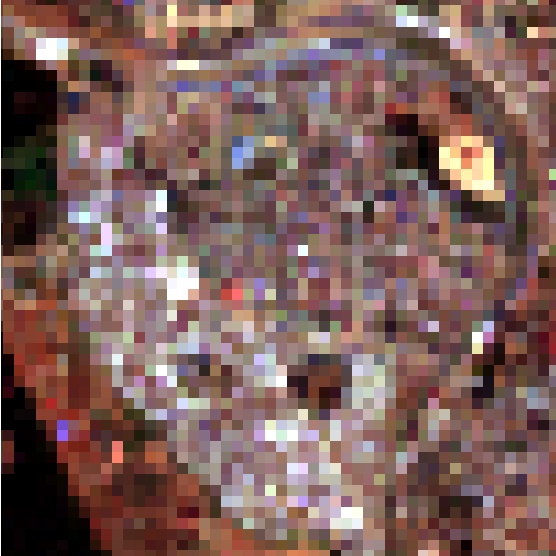}
\end{minipage}
\caption{Two images of the same informal settlement in Kibera, representing the difference between VHR and LR imagery. \textit{Left}: A DigitalGlobe 30cm VHR image. \textit{Right}: The Sentinel-2 10m resolution image.}
\label{fig:compresimg}
\end{figure}

However, solving this problem is challenging due to several factors. 1) It requires collaboration among multiple parties: the NGOs, local government, the remote sensing and machine learning communities. 2) The locations and distribution of these informal settlements have yet to be mapped thoroughly on the ground or aerially, as the mapping demands dedicated human and financial resources. This often leads to partially completed, or completely un-annotated data-sets. 3) Informal settlements tend to grow sporadically (both in space and time), which adds an additional layer of complexity.
4) Even though we have access to satellite imagery for the entire globe, much of this raw data is not in a usable format for machine learning frameworks, making it difficult to extract actionable insights at scale~\cite{xie2015transfer}. 
5) There may be no local government structure in a particular settlement, which can inhibit our ability to gather data quickly.
and make it difficult to extract good quality ground truth data, see Section \ref{sec:data}.\\

In order to address these challenges, in this work we propose a semi-automated framework that takes a satellite image, directly extracted in its \textit{raw}-user form and outputs a trained classifier that produces binary maps highlighting the locations of informal settlements.


Our first approach, the cost-effective approach, takes advantage of the pixel level contextual information by training a classifier to learn a unique spectral signal for informal settlements. 
When we require finer grained features, such as the roof size, or the density of the surrounding settlements to determine whether or not there exists an informal settlement, we demonstrate a second approach that uses a semantic segmentation neural network to extract these features, the cost-prohibitive solution. See Section~\ref{sec:method}.\\

To ensure that this work can be applied in the field, we have had an active partnership with UNICEF, to understand what we can do to facilitate their needs further and how we can facilitate the needs of other NGOs. Because of this, we focused on developing a system that will work in a computationally efficient and monetary effective manner. Our main approach runs efficiently on a laptop, or desktop CPU and is cost-effective as we only use freely available, openly accessible LR satellite imagery, rather than VHR imagery which can cost hundreds-of-thousands of dollars.

\noindent Within this paper we make the following \textbf{contributions}:
\begin{quote}
\begin{itemize}
\item We introduce and extensively validate two machine learning based approaches to detect and map informal settlements. One is cost-effective, the other is cost-prohibitive, but is required when contextual information is needed.
\item We demonstrate for the first-time that informal settlements can be detected effectively using only freely and openly accessible LR satellite imagery.
\item We release to the public two informal settlement benchmarks for LR and VHR satellite imagery, with accompanying ground truths.
\item We provide all source code and models. 
\end{itemize}
\end{quote}
In Section~\ref{sec:data} we provide details of the data used and the challenges involved in collecting it. In Section~\ref{sec:relatedwork} we provide a condensed overview of related work and current approaches. In Section~\ref{sec:method} we introduce details of our methodologies and present the results of our contributions in Section~\ref{sec:results}. Finally we conclude and present future work in Section~\ref{sec:conclusions}.

\section{Data Acquisition}
\label{sec:data}


In this work we use a combination of satellite imagery and on-the-ground measurements. However, to take advantage of machine learning frameworks we require an absolute \emph{ground truth}, which facilitates robust training and validation. Ground truth data for this project was very sparse, in part due to the difficulties and financial costs in obtaining the data across vast regions of developing nations. This meant that much of the accessible data was incomplete. Even when the data was available, it was not necessarily in a workable format; either it was provided as part of a PDF, with no external meta-data, or it was simply in an inaccessible format. As part of this work we fused these data sets together, to generate usable data sets that can be used by the community for developing new machine learning models. Data sets can be found here: \url{https://frontierdevelopmentlab.github.io/informal-settlements/}.

\subsection{Satellite Data} 

In the last ten-years there has been an exponential increase in the number of satellites being launched due to the increase in commercial interests. This has accelerated the amount of satellite imagery available and continues to lower the cost of gaining access to VHR data. However, VHR imagery can still cost hundreds, to thousands, to hundreds-of-thousands of dollars per image, or collection of images and is typically only available through commercial providers. Institutions such as the National Aeronautics and Space Administration~(NASA) and ESA do provide a multitude of freely available multi-spectral imagery, but this is typically of a much lower resolution, approximately $10-20 m$ resolution per pixel, and many of the fine grained features are blurred, see Figure \ref{fig:compresimg}.  This makes it difficult to use a deep learning approach effectively to extract optical features that would be required for distinguishing informal and formal settlements, whereas the VHR imagery, less than $1 m$ resolution per pixel, enables us to do this, especially when we require contextual information, Section~\ref{sec:method}. 
\subsubsection{Sentinel-2}
\begin{figure}[t!]
\centering
\includegraphics[width=0.45\textwidth, height=3cm]{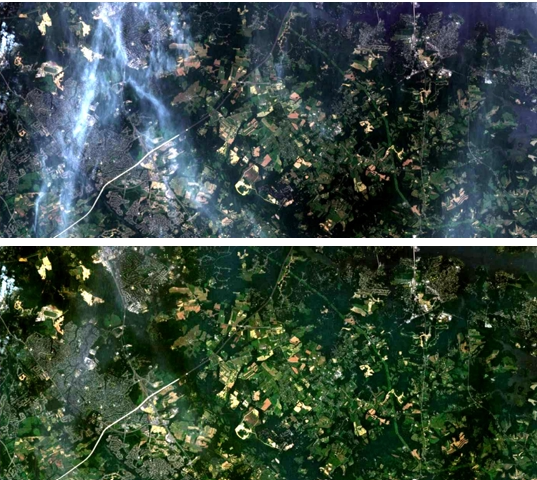}
\caption{Image provided by the~\protect\citeauthor{Esas2img}.\textit{Top}: Represents the Sentinel-2 Level-1C uncorrected image. \textit{Bottom}: Represents the Sentinel-2 Level-2A corrected image. This lower image requires an additional time-consuming computational step to correct for atmospheric distortions in the spectral data. Our method does not require the use of this pre-processing step.}
\label{fig:s2l1cl2a}
\end{figure}

The Sentinel-2 mission is part of the Copernicus program by the European Commission (EC). A global earth observation service addressing six thematic areas: land, marine, atmosphere, climate change, emergency management and security through its Sentinel missions. ESA is responsible for the observation infrastructure of the Sentinels~\cite{Copernicussent2}. The data provided by the Sentinels has a free and open data policy implying that the data from the Sentinel missions is available free of charge to everyone. The ease of data access and use, allows all users from the public, private or research communities to reap the socio-economic benefits of such data~\cite{Waitr2016}.
A Sentinel-2 image is provided to the end user at Level-1C~\cite{sent2userhandbook} and has already gone through a series of pre-processing steps before it reaches the end user. However, these images have not been corrected for atmospheric distortions. This correction requires additional processing time to convert the image into Level-2A product, resulting in bottom of the atmosphere reflectances, see Figure~\ref{fig:s2l1cl2a} for a comparison. 
Within this work we directly use the Level-1C images for our computationally and cost efficient approach, mitigating the need to do the computationally costly processing.

\paragraph{Multi-spectral Data}
The Sentinel-2 satellites map the entire global land mass every 5-days at various resolutions of 10 to 60$m$ per pixel, which means that each pixel represents an area of between $10m^{2}$ to $60m^{2}$. At each resolution, spectral information at the top of the atmosphere (TOA) is provided, creating a total of 13 spectral bands covering the visible, near infrared (NIR) and the shortwave infrared (SWIR) part of the electromagnetic spectrum~\cite{sent2userhandbook,Zhang2017,Drusch2012}. Although there are 13 spectral bands in total, we exclude bands 1, 9 and 10 as they interfere strongly with the atmosphere due to their 60$m$ resolution. This means that we only use the bands 2, 3, 4, 5, 6, 7, 8, 8A, 11, 12 as these bands have minimal interactions with the atmosphere and are provided at either a 10$m$ or 20$m$ spatial resolution.\\

\subsubsection{Very-High-Resolution Satellite Images}
In addition to freely available multi-spectral LR satellite images, we use VHR images with a resolution of up to 30cm per pixel, kindly provided by DigitalGlobe through Satellite Applications Catapult. See Figure~\ref{fig:compresimg} to see the difference in resolution between Sentinel-2 and VHR imagery.We emphasize that VHR imagery is only used in the cost-prohibitive method.

\subsection{Annotated Satellite Imagery}
\begin{figure}[t!]
\centering
\includegraphics[width=0.45\linewidth,height=7cm]{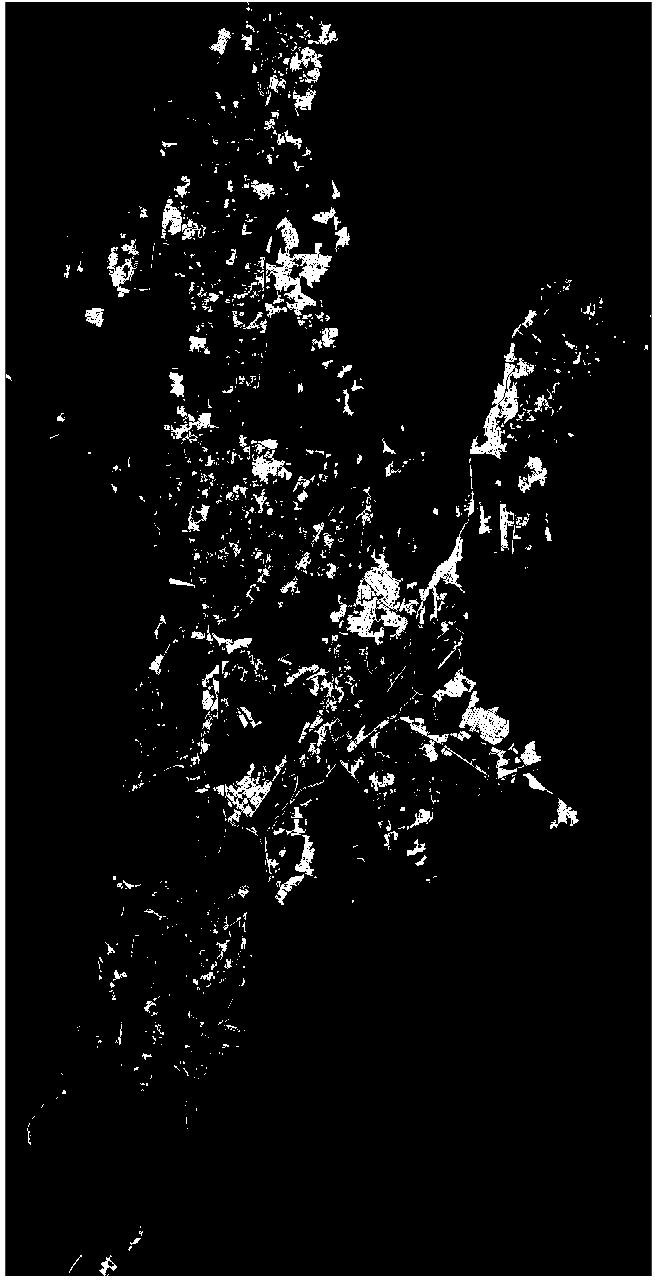}
\includegraphics[width=0.45\linewidth,height=7cm]{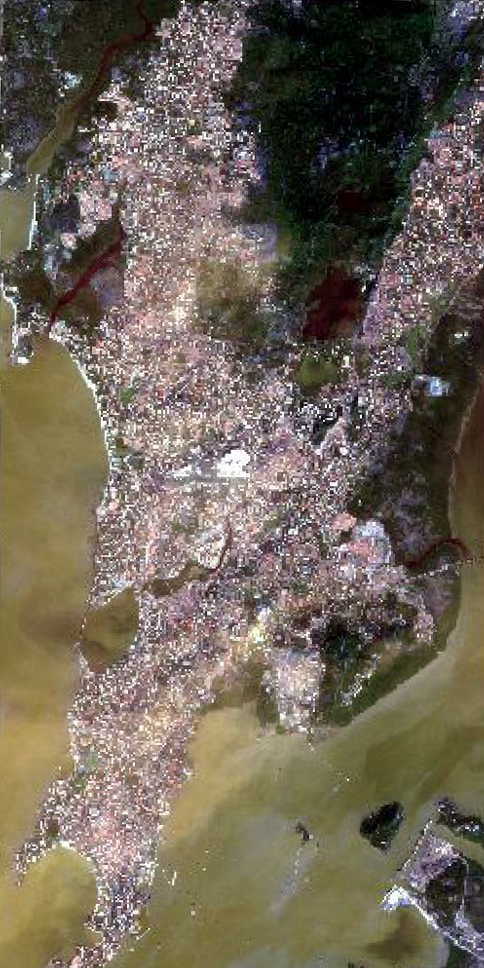}
\caption{An example of annotated ground truth map. \textit{Left:} The city is Mumbai, the white dots represent informal settlements and the black dots represent the environment. \textit{Right:} The Sentinel-2 image of Mumbai.}
\label{fig:mumbaiannotated}
\end{figure}

We have annotated satellite imagery for the locations of informal settlements in parts of  Kenya, South Africa, Nigeria, Sudan, Colombia and Mumbai. We then project these masks on to the satellite image and extract the necessary spectral information at those specific points, see Figure~\ref{fig:mumbaiannotated} for a example of an annotated ground truth map. We have open sourced the necessary code to do this here: 
\url{https://frontierdevelopmentlab.github.io/informal-settlements/}.

\section{Related Work}
\label{sec:relatedwork}
Recent publications applying machine learning to remote sensing data, in particular to satellite imagery, that have focused on detecting, or mapping informal settlements \cite{xie2015transfer,varshney2015targeting,mboga2017detection,mahabir2018critical,kuffer2016slums,ASMAT2012650,Kohli2016} have typically been trained on a specific region, or feature in combination with VHR \cite{4610942,gevaert2016classification,4610942,kuffer2016slums}. The approaches most in spirit to our own are~\cite{varshney2015targeting,xie2015transfer,Jean790}.
\citeauthor{varshney2015targeting} focus on detecting roofs in Eastern Africa using a template matching algorithm and random forest, they take advantage of Google Earths' API to extract high resolution imagery, which although is free to researchers, is not openly available to everyone. \citeauthor{xie2015transfer} and \citeauthor{Jean790} use a mixture of data sources and transfer learning across different data sets to generate poverty maps by taking advantage of night time imagery through the National Oceanic and Atmospheric Administration
(NOAA) and daytime imagery through Google Earths' API. However, to our knowledge there exists no previous work on predicting informal settlements solely from LR data, or predicting informal settlements in the way that we present here. This inhibits our ability to benchmark against previous methods. Thus, by providing the data sets and the baselines in this paper, we provide a robust way to compare the effectiveness of any future approaches and facilitate the creation of new machine learning methodologies.

\section{Methods}
\label{sec:method}

In this section, we describe our approaches for detecting and mapping informal settlements. We introduce two different methods; a cost-efficient method and cost-prohibitive method. Our \textbf{first method} trains a classifier to learn what the spectrum of an informal settlement is, using LR freely available Sentinel-2 data. To do this, we employ a pixel-wise classification, where the system learns whether or not a 10-band spectra is associated to an informal settlement or the environment, which encompasses everything that is not an informal settlement.  
Our \textbf{second method}, is a semantic segmentation deep neural network that uses VHR satellite imagery, which is useful when informal settlements do not have unique spectra when compared to the environment, like those in Sudan, see Figure~\ref{fig:sudaninffor}.
\subsection{Cost Effective Method}

\textbf{Canonical Correlation Forests}~(CCFs)~\cite{rainforth2015canonical} are a decision tree ensemble method for classification and regression.
CCFs are the state-of-the-art random forest technique, which have shown to achieve remarkable results for numerous regression and classification tasks~\cite{rainforth2015canonical}.
Individual canonical correlation trees are binary decision trees with hyper-plane splits based on local canonical correlation coefficients calculated during training. Like most random forest based approaches, CCFs have very few hyper-parameters to tune and typically provide very good performance out of the box. All that has to be set is the number of trees, $n_{trees}$. For CCFs, setting $n_{trees} = 15$ provides a performance that is empirically equivalent to a random forest that has $n_{trees} = 500$~\cite{rainforth2015canonical}, meaning CCFs have lower computational costs, whilst providing better classification.
CCFs work by using canonical correlation analysis~(CCA) and projection bootstrapping during the training of each tree, which projects the data into a space that maximally correlates the inputs with the outputs. This is particularly useful when we have small data-sets, like in our case, as it reduces the amount of artificial randomness required to be added during the tree training procedure and improves the ensemble predictive performance~\cite{rainforth2015canonical}.\\

 The computational efficiency aspects of CCFs and their suitability to both small and large data-sets, makes them ideal for detecting informal settlements for three reasons. First, many of the organisations that we aim to help will not have access to a large amount of compute resources, therefore computational efficiency is important. Second, to run the CCFs for both training and prediction, all that has to be called is one function. This ensures that the end user does not need to be an expert in ensemble methods and makes the method akin to plug and play. Finally, some of our ground truth data sets are relatively small, which means that we must use the data as efficiently as possible, which CCFs allow us to do. 
When VHR and computational cost are not a restriction we can employ a deep learning approach using convolution neural networks~(CNN) to detect informal settlements.

\subsection{Cost Prohibitive Method}

\begin{figure}
\begin{minipage}{.25\textwidth}
  \includegraphics[width=0.88\linewidth]{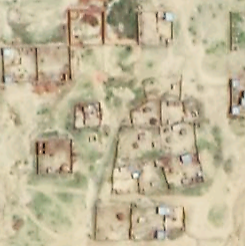}
\end{minipage}%
\begin{minipage}{.25\textwidth}
  \includegraphics[width=0.88\linewidth]{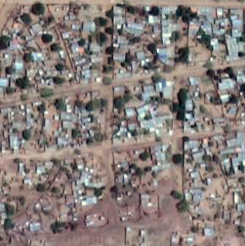}
\end{minipage}
\caption{A VHR image comparing an informal, \textit{left} and formal settlement, \textit{right}, in Al Geneina, Sudan. The main distinguishing feature is the wider contextual information, as the material spectrum's are the same.}
\label{fig:sudaninffor}
\end{figure}

Since informal settlements can also be classified by the rooftop size and the surrounding building density, we employ a state-of-the-art semantic segmentation neural network on optical~(RGB) VHR satellite imagery to detect these contextual features. These contextual features are important when it is not possible to distinguish informal settlements from the environment by spectral signal in the same region. An example of such an informal settlement is shown in Figure~\ref{fig:sudaninffor}. We see that the informal settlements in a rural region of Al Geneina, Sudan have a very low building density, and also the roof tops of both formal and informal settlements are built out of concrete, meaning they have the same spectral signal. This is in contrast to the dense slums in Nairobi and Mumbai.


 \subsubsection{Encoder-Decoder with Atrous Separable Convolution}
\begin{figure*}[ht!]
\centering
\begin{minipage}[b]{0.33\textwidth}
   \includegraphics[height=2.45cm, keepaspectratio]{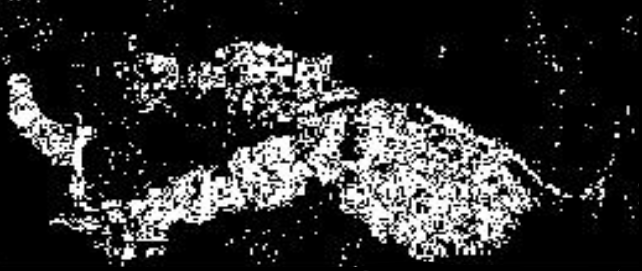}
\end{minipage}
\begin{minipage}[b]{0.33\textwidth}
   \includegraphics[height=2.45cm, keepaspectratio]{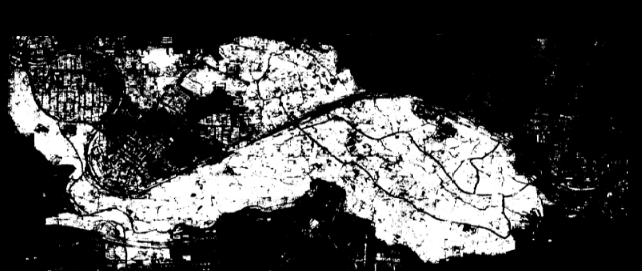}
\end{minipage}
\begin{minipage}[b]{0.33\textwidth}
   \includegraphics[height=2.45cm, keepaspectratio]{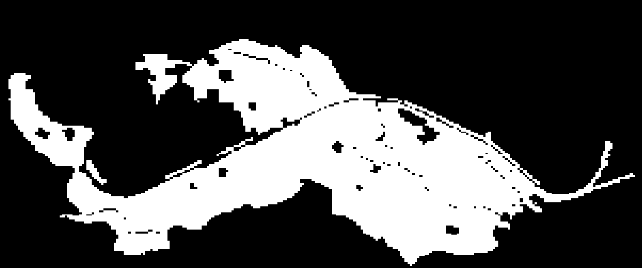}
\end{minipage}

\caption{Predictions of informal settlements (white pixels) in Kibera, Nairobi. \textit{Left:} The CCF prediction of informal settlements in Kibera on low-resolution Sentinel-2 spectral imagery. \textit{Middle:} Deep learning based prediction of informal settlements in Kibera, trained on VHR imagery. \textit{Right: }The ground truth informal settlement mask for Kibera.}
\label{fig:compDLandCCF}
\end{figure*}
For the task of semantic segmentation of informal settlements we use the DeepLabv3+ encoder-decoder architecture. DeepLabv3+ ~\cite{deeplabv3plus2018} is a deep CNN that extends the prior DeepLabv3 network~\cite{chen2017rethinking} with a decoder module to refine the segmentation results of the previous encoder-decoder architecture particularly at the object boarders. The DeepLabv3 architecture uses Atrous Spatial Pyramid Pooling~(ASPP) with Atrous convolutions to explicitly control the resolution at which feature responses are computed within the CNN. 
This ASPP module is augmented with image level features to capture longer range information. 
We use a Xception 65 network backbone in the encoder-decoder architecture. The beneficial use of this Xception model together with applying depth wise separable convolution to ASPP and the decoder modules have been shown in~\cite{deeplabv3plus2018}. 

\subsubsection{Implementation details}

We train the entire network end-to-end with the usual back-propagation algorithm using
eight Tesla V100 GPUs with 16 GBs of memory each.
We initialize the layer weights using those from the pre-trained PASCAL VOC 2012 model~\cite{pascal-voc-2012}. We then fine-tune in turn
the finer strides on the training/validation data.
We train our deep network with a batch size of 32, an initial learning rate of 0.001 and a learning rate decay factor of 0.1 every 2.000 steps until convergence. Our experiments are based on a single-scale evaluation. All other hyper-parameters are the same as in the
DeepLabv3+ model~\cite{deeplabv3plus2018}.

\section{Results}
\label{sec:results}


\paragraph{Experimental Setup} For each region we have a 10-20$m$ resolution Sentinel-2 image, the corresponding VHR 30-50$cm$ resolution image and the ground truth annotations. We have ensured that the images and annotations are aligned in space and time to reduce any additional noise in the data. When training and validating a model on the same region we use a 80-20 split. We ensure that each class contains the same number of points, we then randomly sample 80\% of each class to generate the training data and then use the remaining 20\% of each class to construct our test set, which is comprised of a different set of points. We then center the training data (testing data accordingly) to have a mean of zero and standard deviation of one. We set the $n_{trees} =10$ for training the CCF.
For validating our methods we report both pixel accuracy, and mean intersection over union~(IoU). We use the standard definition of mean IOU, $ meanIOU = \frac{1}{n_{class}}\frac{t_{ii}}{(t_{i} + \sum_j n_{ji} - n_{ii})} $ and pixel classification, $pxclass = \frac{\sum_i n_{ii}}{t_i}$, where $n_{class}$ is the total number of classes, $n_{ij}$ is the number of pixels of class $i$ predicted to belong to class $j$, and $t_i$ is the total number of pixels of class $i$ in ground truth segmentation.\\

We provide a comparison of both the pixel-wise classification with CCFs and the contextual classification with CNNs for the detection and mapping of informal settlements, see Table~\ref{tab:res2}. The CCFs trained solely on freely available and easily accessible low-resolution data perform well, although they are unable to match the performance of the CNN trained on VHR imagery, except for Kibera. Figure~\ref{fig:compDLandCCF} shows the predictions of both methods and the ground truth annotations.  Despite having access to very high resolution data, the CNN still manages to miss-classify structural elements of the informal settlements in Kibera. Whereas the CCF, although more granular, incorporates the full structure of the informal settlement in Kibera via only the spectral information.

\subsubsection{Generalizability} To demonstrate the adaptability of our approach we train each model on different parts of the world and use that model to perform predictions on other unseen regions across the globe. For this paper we train two models, one on Northern Nairobi, Kenya and another on Medellin, Colombia. The results can be found in Table~\ref{tab:res3}. Even though we only have a small amount of data, we are able to demonstrate that our models can generalize moderately well, even with data that is noisy and partially incomplete. We provide several more results in the appendix~\ref{sec:all} of this paper.
 

\begin{table}[h!]
\centering
\begin{tabular}{lcc|cc}
\toprule
 & \multicolumn{2}{c}{\textbf{Pixel Acc.}} & \multicolumn{2}{c}{\textbf{Mean IOU}} \\ \midrule
\textbf{Region} & \textbf{CCF} & \textbf{CNN} & \textbf{CCF} & \textbf{CNN} \\ \midrule
Kenya, Northern Nairobi & 69.4 & 93.1 & 62.0 & 80.8 \\
Kenya, Kibera & 69.0 & 78.2 & 73.3 & 65.5 \\
South Africa, Capetown* & 92.0 & - & 33.2 & - \\
Sudan, El Daien & 78.0 & 86.0 & 61.3 & 73.4 \\
Sudan, Al Geneina & 83.2 & 89.2 & 35.7 & 76.3 \\
Nigeria, Makoko* & 76.2 & 87.4 & 59.9 & 74.0 \\
Colombia, Medellin* & 84.2 & 95.3 & 74.0 & 83.0 \\
India, Mumbai* & 97.0 & - & 40.3 & -
\end{tabular}
\caption{Pixel accuracy and mean IOU (\%) results for informal settlement detection using the CCF pixel-wise classification and the contextual classification with CNNs. CCFs are trained and tested on low resolution imagery, CNNs are trained and tested on VHR imagery. *Represents that the ground truth annotations are less than 75\% complete for the region.}
\label{tab:res2}
\end{table}

\begin{table}[h!]
\centering
\begin{tabular}{lcc|cc}
\toprule
 & \multicolumn{2}{c}{\textbf{Pixel Acc.}} & \multicolumn{2}{c}{\textbf{Mean IOU}} \\ \midrule
\textbf{Region} & \textbf{NN} & \textbf{M} & \textbf{NN} & \textbf{M} \\ \midrule
Kenya, Northern Nairobi & 69.4 & 55.0 & 62.0 & 54.4 \\
Kenya, Kibera & 67.3  & 63.8 & 54.1 & 56.0\\
South Africa, Capetown* & 41.3 & 71.5 & 43.1 & 32.0 \\
Sudan, El Daien & 14.2 & 1.1 & 37.9 & 34.0 \\
Sudan, Al Geneina & 27.1 & 6.0 & 34.9 & 41.0 \\
Nigeria, Makoko* & 59.0 & 77.0 & 37.8 & 34.6 \\
Colombia, Medellin* & 65.0 & 84.2 & 46.9 & 74.0 \\
India, Mumbai* & 37.9 & 63.0 & 32.4 & 34.4
\end{tabular}
\caption{Pixel accuracy and mean IOU (\%) results for informal settlement detection using pixel-wise classification with CCFs trained on a particular region and testing on all other regions. Results are for a model trained on Northern Nairobi~(NN) and a model trained on Medellin~(M). * Represents that the ground truth annotations are less than 75\% complete for the region.}
\label{tab:res3}
\end{table}

\newpage
\section{Conclusions and Future Work}
\label{sec:conclusions}
\paragraph{Conclusion}

In this work we have composed a series of annotated ground truth data-sets and have provided for the first time benchmarks for detecting informal settlements. We have provided a comprehensive list of the challenges faced in mapping informal settlements and some of the constraints faced by NGOs. In addition to this, we have proposed two different methods for detecting informal settlements, one a cost-effective method, the other a cost-prohibitive method. The first method used computationally efficient CCFs to learn the spectral signal of informal settlements from LR satellite imagery. The second used a CNN combined with VHR satellite imagery to extract finer grained features. We extensively evaluated the proposed methods and demonstrated the generalization capabilities of our methods to detect informal settlements not just in a local region, but globally. In particular, we demonstrated for the first time that informal settlements can be detected effectively using only freely and openly accessible multi-spectral low-resolution satellite imagery.

\paragraph{Future work} 
Because of the uncertainties within the ground truth annotations and the differences in informal settlements across the world, we believe that this problem would be very useful for testing transfer learning and meta-learning approaches. In addition to this, Bayesian approaches would enable us to characterize these uncertainties via probabilistic models. This would provide an effective way to create adaptable models that learn what it means for an informal settlement to be informal, as the model absorbs new information.

It is also interesting to note that a 1~$km^{2}$ area containing informal settlements could house up to 129089 people~\cite{desgroppes2011kibera} and so each pixel could represent up to 13 people~\footnote{$100m^{2} = 0.00001km^{2}$, therefore $129089 \times 0.00001 \sim 13 $ people per pixel~($ppx^{-1}$).}. This therefore allows us to also add population estimates to our maps, which UNICEF state is also crucial. This would enable governments and NGOs to understand how much infrastructure is required and how much aid needs to be provided. Although we could have added population estimates in this current work, we have chosen to omit them as it would be irresponsible of us to provide estimates when not enough ground truth data exists, regarding average population numbers in informal settlements. We are actively working with UNICEF to gather more ground truth data for this and additional annotations for informal settlements, as UNICEF would actively like to deploy a system like the one that we have developed here to provide both mapping and population estimates of rural and urban informal settlements.  

\section{Acknowledgements}
This project was executed during the Frontier Development Lab, Europe program, a partnership between the $\phi$-Lab at ESA, the Satellite Applications~(SA) Catapult, Nvidia Corporation, Oxford University and Kellogg College. We gratefully acknowledge the support of Adrien Muller and Tom Jones of SA Catapult for their useful comments, providing VHR imagery and ground truth annotations for Nairobi. We thank UNICEF, in particular Do-Hyung Kim and Clara Palau Montava, for valuable discussions and AIData for access to geo-located Afrobarometer data. We thank Nvidia for computation resources. We thank Yarin Gal, Adam Golinski and Ben Fernando for their for their helpful comments. Bradley Gram-Hansen was also supported by the UK EPSRC CDT in Autonomous Intelligent Machines and Systems (EP/L015897/1). Patrick Helber was supported by the NVIDIA AI Lab program and the BMBF project DeFuseNN (Grant 01IW17002). 
\newpage
\bibliographystyle{ACM-Reference-Format}
\balance
\bibliography{refs}

\clearpage
\appendix
\section{Informal Settlement Classification Results}
\label{sec:all}
In this section we provide all results generated by the CCF on all informal settlement data-sets at our disposal and provides results in terms of two metrics; pixel-wise classification and mean intersection-over-union. We present two sets of results for Capetown. 
\begin{itemize}
\item Capetown with Ground Truth~(CapetownGT) - This represents around $20km^{2}$ region of Capetown (11\% of the city). This data is annotated and it was used in the main paper.
\item Capetown - This represents the whole of Capetown, however, we only have annotations for the region mentioned above~(CapetownGT).
\end{itemize}
\tabcolsep=0.10cm\begin{table*}[ht!]\small
\centering
\begin{tabular}{l|ccccccccccccc}
\toprule
Country & & Colombia & India & Kenya &  &  & Nigeria & South Africa &  & Sudan &  &  &  \\
City & & Medellin & Mumbai & N. Nairobi & Kibera & Kianda & Mokako & Capetown & CapetownGT & AlGeneina & ElDaein &  &  \\
\midrule
Model & Mumbai &  &  &  &  &  &  &  &  &  &  &  &  \\ \midrule
Informal & & 78.35\ & 97.18\ & 71.44\ & 74.84\ & 89.23\ & 94.81\ & 80.88\ & 89.10\ & 21.00\ & 10.96\ &  &  \\
Environment & & 36.32\ & 95.92\ & 43.37\ & 51.65\ & 31.33\ & 28.53\ & 56.45\ & 43.11\ & 68.19\ & 83.71\ &  &  \\
 &  &  &  &  &  &  &  &  &  &  &  &  \\ \midrule
Model & Capetown & &  &  &  &  &  &  &  &  &  &  &  \\ \midrule
Informal & & 96.86\ & 98.00\ & 99.21\ & 96.52\ & 100.00\ & 99.73\ & 98.77\ & 99.61\ & 99.99\ & 99.90\ &  &  \\
Environment &  & 9.80\ & 46.97\ & 9.07\ & 10.12\ & 4.03\ & 7.27\ & 96.33\ & 15.22\ & 0.23\ & 0.10\ &  &  \\
 &  &  &  &  &  &  &  &  &  &  &  &  \\ \midrule
Model & CapetownGT & &  &  &  &  &  &  &  &  &  &  &  \\  \midrule
Informal &  & 83.77\ & 73.94\ & 93.59\ & 82.16\ & 96.13\ & 95.32\ & 84.78\ & 92.34\ & 99.48\ & 90.82\ &  &  \\
Environment & & 51.69\ & 85.41\ & 40.25\ & 45.33\ & 22.23\ & 24.60\ & 82.68\ & 93.33\ & 2.53\ & 5.99\ &  &  \\
 &  &  &  &  &  &  &  &  &  &  &  &  \\ \midrule
Model & N. Nairobi & &  &  &  &  &  &  &  &  &  &  &  \\  \midrule
Informal &  & 64.65\ & 37.91\ & 69.45\ & 67.63\ & 90.33\ & 58.55\ & 36.06\ & 41.33\ & 27.07\ & 14.18\ &  &  \\
Environment & & 63.64\ & 62.44\ & 69.27\ & 73.44\ & 50.75\ & 53.85\ & 75.21\ & 81.57\ & 68.77\ & 89.17\ &  &  \\
 &  &  &  &  &  &  &  &  &  &  &  &  \\ \midrule
Model & Kibera & &  &  &  &  &  &  &  &  &  &  &  \\  \midrule
Informal &  & 48.23\ & 40.82\ & 63.65\ & 67.84\ & 79.56\ & 57.71\ & 46.50\ & 45.07\ & 30.96\ & 31.31\ &  &  \\
Environment & & 72.04\ & 27.18\ & 76.56\ & 75.56\ & 67.73\ & 58.38\ & 54.76\ & 55.92\ & 62.13\ & 65.53\ &  &  \\
 &  &  &  &  &  &  &  &  &  &  &  &  \\ \midrule
Model & Kianda & &  &  &  &  &  &  &  &  &  &  &  \\  \midrule
Informal & & 45.23\ & 43.18\ & 62.57\ & 61.09\ & 90.97\ & 46.64\ & 24.45\ & 26.79\ & 68.75\ & 53.73\ &  &  \\
Environment &  & 82.06\ & 60.65\ & 85.33\ & 83.49\ & 75.48\ & 62.19\ & 77.23\ & 86.27\ & 36.50\ & 57.57\ &  &  \\
 &  &  &  &  &  &  &  &  &  &  &  &  \\ \midrule
Model & AlGeneina & &  &  &  &  &  &  &  &  &  &  &  \\ \midrule
Informal & & 81.21\ & 66.80\ & 80.73\ & 79.76\ & 85.36\ & 56.07\ & 52.94\ & 60.52\ & 83.17\ & 94.68\ &  &  \\
Environment &  & 46.79\ & 48.97\ & 49.71\ & 31.87\ & 17.26\ & 38.46\ & 76.21\ & 63.86\ & 79.43\ & 5.25\ &  &  \\
 &  &  &  &  &  &  &  &  &  &  &  &  \\ \midrule
Model & ElDaein & &  &  &  &  &  &  &  &  &  &  &  \\ \midrule
Informal & & 24.07\ & 9.50\ & 46.58\ & 54.11\ & 71.82\ & 8.37\ & 29.90\ & 25.65\ & 68.52\ & 78.16\ &  &  \\
Environment & & 76.61\ & 40.11\ & 67.18\ & 71.47\ & 71.20\ & 85.19\ & 48.91\ & 61.22\ & 30.61\ & 71.65\ &  &  \\
 &  &  &  &  &  &  &  &  &  &  &  &  \\ \midrule
Model & Mokako & &  &  &  &  &  &  &  &  &  &  &  \\ \midrule
Informal & & 17.33\ & 69.43\ & 10.67\ & 14.15\ & 13.54\ & 76.27\ & 35.60\ & 34.76\ & 5.53\ & 10.57\ &  &  \\
Environment & & 85.22\ & 39.06\ & 89.91\ & 94.07\ & 94.37\ & 80.84\ & 60.91\ & 61.89\ & 92.97\ & 88.51\ &  &  \\
 &  &  &  &  &  &  &  &  &  &  &  &  \\ \midrule
Model & Medellin & &  &  &  &  &  &  &  &  &  &  &  \\  \midrule
Informal & &  84.06\ & 62.95\ & 54.10\ & 63.84\ & 88.12\ & 77.00\ & 63.38\ & 71.46\ & 5.66\ & 1.09\ &  &  \\
Environment & & 79.88\ & 63.02\ & 85.79\ & 78.85\ & 47.00\ & 37.45\ & 62.01\ & 57.68\ & 96.28\ & 95.78\ &  &  \\ \midrule
\end{tabular}
\caption{Pixel classification scores. The \textit{Model} row represents the model that has been trained on the stated city and the\\ columns represent the city on which that trained model is making a prediction on.}
\end{table*}

 \begin{table*}[ht!]\small
 \centering
 \begin{tabular}{l|cccccccccccc}
 \toprule
 Country & & Colombia & India & Kenya &  &  & Nigeria & South Africa &  & Sudan &  &  \\
 City & & Medellin & Mumbai & N. Nairobi & Kibera & Kianda & Mokako & Capetown & CapetownGT & AlGeneina & ElDaein &  \\ \midrule
 Model & Mumbai & &  &  &  &  &  &  &  &  &  &  \\ \midrule
 Informal  IOU & &  31.23 & 68.58 & 41.42 & 45.04 & 30.14 & 27.74 & 56.32 & 42.83 & 55.82 & 59.81 &  \\
 Environment IOU & & 42.45 & 12.17 & 16.08 & 40.91 & 29.50 & 41.21 & 1.82 & 6.15 & 9.84 & 8.04 &  \\
 Mean IOU & & 36.84 & 40.37 & 28.75 & 42.97 & 29.82 & 34.47 & 29.07 & 24.49 & 32.83 & 33.92 &  \\
  &  &  &  &  &  &  &  &  &  &  &  \\ \midrule
 Model & Capetown & &  &  &  &  &  &  &  &  &  &  \\ \midrule
 Informal  IOU & & 9.57 & 46.93 & 9.06 & 9.92 & 4.03 & 7.26 & 32.00 & 15.22 & 0.23 & 0.09 &  \\ 
 Environment IOU & & 44.07 & 7.53 & 15.20 & 37.97 & 26.14 & 37.11 & 1.44 & 4.74 & 21.95 & 30.96 &  \\
 Mean IOU & & 26.82 & 27.23 & 12.13 & 23.94 & 15.09 & 22.18 & 16.72 & 9.98 & 11.09 & 15.53 &  \\
  &  &  &  &  &  &  &  &  &  &  &  \\ \midrule
 Model & CapetownGT & &  &  &  &  &  &  &  &  &  &  \\ \midrule
 Informal  IOU & & 46.06 & 84.43 & 39.83 & 40.97 & 21.94 & 23.98 & 82.54 & 57.74 & 2.53 & 5.76 &  \\
 Environment IOU & & 51.03 & 17.16 & 20.20 & 42.36 & 29.22 & 40.19 & 4.60 & 8.81 & 22.33 & 29.34 &  \\
 Mean IOU & & 48.55 & 50.80 & 30.02 & 41.67 & 25.58 & 32.08 & 43.57 & 33.28 & 12.38 & 17.55 &  \\
  &  &  &  &  &  &  &  &  &  &  &  \\ \midrule
 Model & N. Nairobi &  &  &  &  &  &  &  &  &  &  \\ \midrule
 Informal  IOU & & 50.26 & 60.77 & 80.99 & 61.79 & 49.14 & 43.86 & 74.67 & 78.82 & 57.08 & 64.38 &  \\
 Environment IOU & & 43.60 & 3.98 & 42.15 & 46.46 & 36.87 & 31.83 & 1.41 & 7.44 & 12.81 & 11.43 &  \\
 Mean IOU & & 46.93 & 32.38 & 61.57 & 54.12 & 43.00 & 37.84 & 38.04 & 43.13 & 34.95 & 37.90 &  \\
  &  &  &  &  &  &  &  &  &  &  &  \\ \midrule
 Model & Kibera & &  &  &  &  &  &  &  &  &  &  \\ \midrule
 Informal  IOU & & 51.83 & 26.49 & 72.33 & 78.51 & 63.33 & 47.37 & 54.42 & 54.16 & 52.05 & 50.09 &  \\
 Environment IOU & & 35.17 & 2.33 & 26.25 & 68.11 & 40.79 & 32.84 & 1.03 & 4.15 & 13.18 & 17.71 &  \\
 Mean IOU & & 43.50 & 14.41 & 49.33 & 73.31 & 52.06 & 40.11 & 27.73 & 29.16 & 32.61 & 33.90 &  \\
  &  &  &  &  &  &  &  &  &  &  &  \\ \midrule
 Model & Kianda & &  &  &  &  &  &  &  &  &  &  \\ \midrule
 Informal  IOU & & 58.10 & 59.17 & 80.38 & 68.05 & 68.33 & 48.09 & 76.57 & 82.69 & 33.56 & 47.67 &  \\
 Environment IOU & & 36.53 & 4.35 & 33.07 & 47.61 & 50.43 & 27.63 & 1.06 & 6.25 & 21.07 & 27.62 &  \\
 Mean IOU & &  47.31 & 31.76 & 56.72 & 57.83 & 59.38 & 37.86 & 38.81 & 44.47 & 27.31 & 37.65 &  \\
  &  &  &  &  &  &  &  &  &  &  &  \\ \midrule
 Model & AlGeneina & &  &  &  &  &  &  &  &  &  &  \\ \midrule
 Informal  IOU & & 40.99 & 48.26 & 48.19 & 28.51 & 16.44 & 30.98 & 75.80 & 62.40 & 41.20 & 5.12 &  \\
 Environment IOU & & 47.58 & 5.31 & 19.90 & 36.78 & 24.84 & 26.45 & 2.12 & 6.15 & 30.17 & 30.43 &  \\
 Mean IOU & & 44.29 & 26.79 & 34.04 & 32.64 & 20.64 & 28.71 & 38.96 & 34.28 & 35.69 & 17.78 &  \\
  &  &  &  &  &  &  &  &  &  &  &  \\ \midrule
 Model & ElDaein & &  &  &  &  &  &  &  &  &  &  \\ \midrule
 Informal  IOU & & 48.74 & 38.57 & 61.75 & 56.39 & 64.98 & 56.66 & 48.52 & 58.64 & 28.13 & 68.01 &  \\
 Environment IOU & &  18.36 & 0.65 & 15.55 & 36.33 & 38.86 & 6.59 & 0.58 & 2.56 & 19.73 & 54.78 &  \\
 Mean IOU & & 33.55 & 19.61 & 38.65 & 46.36 & 51.92 & 31.63 & 24.55 & 30.60 & 23.93 & 61.39 &  \\
  &  &  &  &  &  &  &  &  &  &  &  \\ \midrule
 Model & Mokako & &   &  &  &  &  &  &  &  &  &  \\ \midrule
 Informal  IOU & & 52.52 & 38.54 & 78.39 & 62.69 & 72.95 & 63.65 & 60.46 & 59.59 & 73.49 & 63.16 &  \\
 Environment IOU & & 14.48 & 4.69 & 6.61 & 12.84 & 11.61 & 56.05 & 0.90 & 3.55 & 4.42 & 8.42 &  \\
 Mean IOU & & 33.50 & 21.61 & 42.50 & 37.77 & 42.28 & 59.85 & 30.68 & 31.57 & 38.96 & 35.79 &  \\
  &  &  &  &  &  &  &  &  &  &  &  \\ \midrule
 Model & Medellin & &  &  &  &  &  &  &  &  &  &  \\ \midrule
 Informal  IOU & & 75.06 & 62.00 & 79.78 & 65.12 & 45.18 & 33.33 & 61.76 & 56.72 & 76.12 & 66.33 &  \\
 Environment IOU & & 72.73 & 6.71 & 29.03 & 46.84 & 34.41 & 36.01 & 1.64 & 6.46 & 4.99 & 0.99 &  \\
 Mean IOU & & 73.90 & 34.36 & 54.40 & 55.98 & 39.79 & 34.63 & 31.70 & 31.59 & 40.56 & 33.66 &  \\ \midrule
 \end{tabular}
 \caption{Intersection Over Union (IOU) results. The \textbf{Model} row represents the model that has been trained on the stated city and the columns represent the city on which that trained model is making a prediction on.}
 \end{table*}
 \cleardoublepage
 \section{Very High-Resolution vs. Low-Resolution Images}
 \label{sec:largescale}
Here we provide a comparison of differing image qualities, 50cm VHR image provided by DigitalGlobe and 10m LR image freely available via the Sentinel hub, for a subset of Kibera. In comparing the two images it is clear to see why we must use the additional spectral data contained within the LR images, if we are to extract any useful information regarding whether or not that pixel contains a settlement. 

 \begin{figure*}[ht!]
 \includegraphics[width=\linewidth,keepaspectratio]{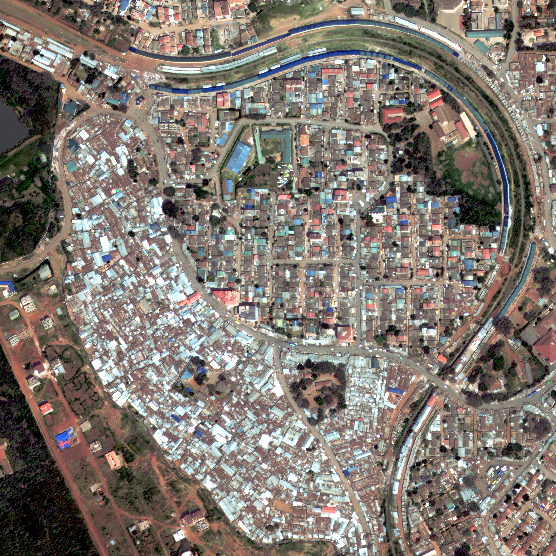}
 \caption{A very high-resolution (VHR) image of the Kibera slum. As depicted in Figures 2 and 5 in the main paper. A full resolution excerpt of this VHR satellite imagery is shown in Figure~\ref{fig:kibera_slum_excerpt_full_resolution}.}
 \end{figure*}

 \begin{figure*}[ht!]
 \includegraphics[width=\linewidth,keepaspectratio]{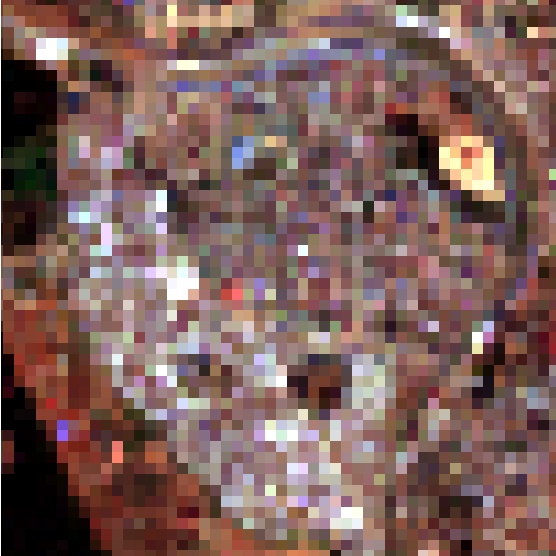}
 \caption{A low-resolution Sentinel-2 image of the Kibera slum as used in this study. As depicted in Figure 2 and 5 from the main paper.}
 \label{fig:kibera_slum_excerpt_full_resolution}
 \end{figure*}

\end{document}